**Einstein's quantum clocks and Poincaré's classical clocks
in special relativity**

Yves Pierseaux
Faculté des Sciences physiques, Fondation Wiener-Anspach, Université Libre de Bruxelles ,
ypiersea@ulb.ac.be
Subfaculty of Philosophy, University of Oxford,
yves.pierseaux@philosophy.ox.ac.uk

# 1 Principles of relativity and group structure of LT

The interest to learn the two approaches of special relativity (SR) has been notably emphasised by John Bell in "How to teach special relativity" (1987):

It is my impression that those [students] with a more classical education [including Fitzgerald contraction], knowing something of the reasoning of Larmor, Lorentz and Poincaré, as well that of Einstein, have stronger and sounder instincts.[13]

According to Bell there is, between Einstein's and Lorentz-Poincaré's reasoning, only a "difference of philosophy and a difference of style". The first question here considered is to know if there exist only two approaches (two interpretations whose one implies a more classical education) of SR or two genuine theories of SR?

An essential element of a theory of SR is naturally the formulation of a principle of relativity. Lorentz's paper of 1904 "Electromagnetic phenomena in a system moving with any velocity less than that of light"[3] is not based on a principle of relativity[1]. Contrary to Lorentz's approach [1], Einstein's and Poincaré's works are based on a principle of relativity:

| **Henri Poincaré** | **Albert Einstein** |
|---|---|
| "La Dynamique de l'électron" [9-9b] (5 June - 23 July 1905) | "Zur Electrodynamik bewegter Körper" [3] (27 June 1905) |
| This impossibility of experimentally demonstrating the absolute motion of the Earth appears to be a general law of the Nature; it is reasonable to assume existence of this law, which we shall call the relativity postulate, and to assume that it is universally valid. [9b, introduction, 1905] | 1° The laws by which the states of physical systems undergo change are not affected, whether these changes of state (Zustandänderungen) be referred to the one or the other systems of two systems of co-ordinates in uniform translatory motion [3, §2,1905] |

---

[1]Lorentz admitted [16] that his point of view was not relativistic and that Einstein and Poincaré had a relativistic point of view. The historical Poincaré's mistake (perhaps by excess of modesty) was certainly to have called his own principle of relativity the "Lorentz's principle of relativity" (1911).

Both 1905 works are almost simultaneous and largely independent. We don't insist here on the formulation of the invariant laws, "les lois du milieu électromagnétique" for Poincaré and "Die Gesetze diese Zustandsveränderungen" for Einstein.

We insist on the fact that *nowhere*, in Poincaré's work on SR (from 1900 to 1912) [6 to 12], the invariance of the speed of light (second Einstein's principle) appears to be a basic principle.

A second essential element for a theory of SR is of course Lorentz's transformations (LT, in the following of this paper, I adopt respective notations of both authors):

**Poincaré's LT**

The essential idea of Lorentz consists in that the equations of the electromagnetic field will not be altered by a certain transformation (which I shall further term the Lorentz transformation) of the following form
$x' = k\, l\, (x - \varepsilon t)$, $y' = l\, y$, $z' = l\, z$, $t' = k\, l\, (t - \varepsilon x)$
where x, y, z, are the co-ordinates and t the time before the transformation, and x', y', z', and t, are the same after the transformation. [9b, §1, 1905]

**Einstein's LT**

To any system of values x, y, z, t, which completely defines the place and the time on an event in the stationary system K, there belongs a system of values $\xi, \eta, \zeta, \tau$, determining that event relatively to the system k, and our task is now to find the system of equations connecting this quantities (…).
We obtain: $x = \varphi\, \gamma\, (x-vt)$, $\eta = \varphi\, y$, $\zeta = \varphi\, z$, $\tau = \varphi\, \gamma\, (t - vx/c^2)$ [3, §3, 1905]

The transformations of co-ordinates of an event (I don't insist here on the crucial role of Einstein's concept of event) are deduced by Einstein in the kinematics part of his article while they are induced by Poincaré from the covariance of Maxwell-Lorentz's equations[2].

A third crucial element for a theory of SR is the structure of group of LT. The physicists credit often Einstein with the discovery of the relativistic law of composition of the speed and Poincaré with the discovery of the structure of group of LT. But in fact the two elements are in each approach (Einstein's § 5 and Poincaré's § 4).

The interesting point is not in these polemical questions of priority but in the question to know *if we must delete the ether because the LT form a group*. Let us examine this question in details in Poincaré's work.

---

[2] It is however not true that Poincaré's SR would be less general as Einstein's SR because Poincaré applies LT to gravitation force.

Indeed in Lorentz's conception there are two systems and one of them, where the ether is at rest in absolute space, is privileged. This is also the case in §1 of Poincaré's paper. But what happens with the ether if there are three systems K, K', K" connected by three LT of the same form, when Poincaré establishes in his §4 the group structure (transitivity)? :

It is noteworthy that the Lorentz transformations form a group. For, if we put:

$$x' = l\, k\, (x + \varepsilon t), \qquad y' = l\, y, \qquad z' = l\, z, \qquad t' = k\, l\, (t + \varepsilon x)$$

and

$$x'' = l'\, k'(x' + \varepsilon' t'), \qquad y' = l'\, y, \qquad z' = l'\, z, \qquad t'' = l'\, k'\, (t' + \varepsilon' x')$$

we find that

$$x' = l''\, k''\, (x + \varepsilon'' t), \qquad y'' = l''\, y, \qquad z'' = l''\, z, \qquad t'' = l''\, k''\, (t + \varepsilon'' x)$$

with

$$\varepsilon'' = \frac{\varepsilon + \varepsilon'}{1 + \varepsilon \varepsilon'}$$

[9b, §4, 1905]

The group structure is of course totally incompatible with the existence of a privileged system. Indeed, if the three systems are connected (two to two) by the three LT[3], it is *logically* impossible to maintain any absolute conception. One could however object that this property of transitivity is demonstrated by the mathematician Poincaré on the general group of 2 parameters (l, ε) where the parameter *l* has not got any physical meaning. But the "mathematician" Poincaré writes at the end of his §4 that his group with 2 parameters (l, ε) must be reduced to a group with one parameter $l\,(\varepsilon)$ because the only physical concept in question here is the concept of velocity :

For our purposes, however, we have to consider only certain of the transformations in this group. We must regard *l* as being a function of ε, the function being chosen so that this partial group is itself a group. [9b, §4]

Poincaré shows that for this subgroup with one parameter ε, we must have $l\,(\varepsilon) = 1$. He insists in the introduction on the importance of his own demonstration that implies the nature

---

[3] This is clearly the difference with Lorentz's theory in which to pass from one material system for another he uses the Galilean transformation (see Miller, [18])

purely longitudinal of the Lorentz's contraction but also, and above all that ε is a relative velocity. The ether is not deleted but only a relative velocity with respect to it can have got a physical meaning

The conception according to which "the absolute space physically exists but it is impossible to measure an absolute speed with respect to it" is not a Poincaré's conception but a Lorentz's conception. Lorentz's point of view is the starting point of Poincaré (§ 1) but not the final point (§ 4).

Poincaré explains in "La relativité de l'espace"(1907)"this fundamental issue of his 1905 work. [10]: the concept absolute space has not got any physical meaning but only a psychological meaning "Whoever speaks of absolute space uses a word devoid of meaning".

In the same text he dissociates clearly the concept of absolute space from this one of ether ("I mean this time not its absolute velocity, which has no sense, but is velocity in relation with the ether). But the splitting between concepts of absolute and relative velocity with respect to the ether is not clearly developed in this text. On the other hand the concept of relative velocity with respect to the ether is clearly defined, one year later, in "La dynamique de l'electron"(1908):

However that be, it is impossible to escape the impression that the principle of Relativity is a general law of nature, and that we shall never succeed, by any imaginable method in demonstrating any but *relative velocities* and by this I mean not merely the velocities of bodies *in relation to the ether* but the velocities of bodies in relation to each other. [11, §6, 1908]

According to Poincaré the bodies and the ether must be treated exactly of the same manner with respect to the concept of relative velocity. In the same text, Poincaré explains that ether can be regarded *by definition* in (absolute) rest:

It is not a question, of the velocity in relation to absolute space, but the velocity in relation to the ether, which is regarded *by definition*, as being in absolute repose."(in italics in the text).[10]

What is the meaning of in absolute repose *by definition*?
The question was : in which system is the ether at rest ? in K, K', K"?
The answer is not that it is "hidden" but that Poincaré's ether doesn't have got any singular state of movement. In others words in order to study the laws of physics in two different inertial system Poincaré's ether can be chosen *by definition*, for each couple of inertial systems (KK', KK", K'K", at rest in one of the two frames but the other one is then in movement with respect to the ether.

We see so that the difficulty the two theories is mainly logic because on the level of mathematical physics the problem is (almost) completely solved by Poincaré in 1905. There are two logical *relativistic* answers to the negative results of Michelson's experiment. The first consists to closely associate ether and absolute space and to delete the both. This is Einstein's well known answer. The second consists to radically dissociate ether and absolute space and to transform the absolute ether into a relativistic ether.

The situation is very odd because Einstein's SR denies the existence of an ether while Poincaré' SR affirms the existence of a relativistic and deformable ether. There is a really antinomy in the meaning of Kant. We want to transform this philosophical antinomy into a physical opposition between both SR with and without ether.

We want therefore clearly separate our analysis from this one that consists to say that absolute ether is hidden by Poincaré. Poincaré's relativistic ether is not a ghost artificially introduced in Einstein's axiomatic. It exerts an (enormous) pressure on the electron not only in order to balance the electrostatic repulsion but also to contract the deformable electron.

## 2 Poincaré's principles and Einstein's principles of SR

We showed (1) that in each theory there are: a principle of relativity and the structure of group of LT. But one could object that Poincaré's theory is not completed because the kinematics is missing and that his relativistic dynamics rests on only one principle (the principle of relativity). This is true for the first part of the fundamental Poincaré's 1905 work, until his §4. But if we know of course Einstein's second principle, we don't know Poincaré's second principle that is developed in the second part of Poincaré's 1905 work (§4-§9).

Let us firstly examine the historical situation. In 1904 Poincaré, in his conference on "the principles of mathematical physics", just after his first formulation of principle of relativity, Poincaré underlines the necessity to admit another principles:

Unhappily, that does not suffice, and *complementary hypotheses* are necessary. It is necessary to admit that bodies in motion undergo a uniform contraction in the sense of the motion. (my italics) [8, 1904]

If Poincaré put "hypotheses" at the plural it is not because the hypotheses of uniform contraction would not be sufficient but because, already in 1900 [7], he is looking for a dynamical force exerted by the ether on the bodies to justify Lorentz's hypothesis (**LH**) of contraction in order to reconcile Lorentz's theory with the principle of reaction. In his fundamental 1905 work, he determines this force:

But in the Lorentz hypothesis [LH], also, the agreement between the formulas does not occur just by itself; it is obtained together with a possible explanation of the compression of the electron under the assumption that the *deformed and compressed electron is subject to constant external pressure*, the work done by which is proportional to the variation of volume of this electron. (my italics) [9b]

The first part of Poincaré's 1905 work consists to show that LT form group (§ 4) and the second part (§5, 6, 7) to show that a complementary force[4] must be introduce to dynamically not only in order to balance the electrostatic repelling force (as it is generally admitted)[5] but also, and above all, in order to justify LH (real contraction of the electron)[6]. The deformed and compressed electron is subject to constant external pressure of the relativistic and deformable[7] ether (according to the principle of reaction).

According to Poincaré, the principle of relativity and the principle of real contraction are dynamically complementary[8]. It is however true that Poincaré 1905 work is not presented on an axiomatic basis as Einstein's kinematics.

We shall show that there exists a implicit kinematics, underlying Poincaré's relativistic dynamic, that is based on the compatibility between the principle of relativity and principle of real contraction (LH). According to the "fine structure" we must develop respective logic of two great spirits, Poincaré and Einstein. So, in respecting the *spirit* of Poincaré's text and Einstein's text, the opposition rigid-deformable doesn't have to take to the letter but must be understood in that way:

---

[4] Poincaré obtains (§6 of the paper) the fundamental equation of relativistic dynamics (by the using of a principle of least action taking in account "Poincaré's pressure").

[5] The relativistic mechanics of continuous medium is the starting point of Poincaré's SR and the final point of Einstein's SR. Laue in particular rediscovers in 1911 Poincaré's pressure but in a purely static sense while in Poincaré's text this latter has an explicit dynamical sense (with an implicitly kinematics sense).

[6] The relativistic mechanics of continuous medium is the starting point of Poincaré's SR and the final point of Einstein's SR. Laue in particular rediscovers in 1911 (and Fermi ten years later) Poincaré's pressure but in a purely static sense (more exactly: electrostatic meaning) while in Poincaré's text this latter has an explicit dynamical meaning (with an implicitly kinematics meaning we develop here).

[7] The non relativistic ether is of course perfectly rigid. Poincaré's relativistic ether, directly comes from the covariance of all the Maxwell-Lorentz equations (transversal waves), is *deformable*. The non relativistic notion of rigidity is irrelevant as well for Poincaré's ether than for Einstein's rods.

1) There is an underlying kinematics "of Poincaré's deformable rods", based, as Einstein's kinematics, on "fundamental principles".
2) The important concept in Einstein's kinematics is not the rigidity of rods but the identity of the rods within both inertial frames.

Let us develop firstly this second point. A superficial analysis could let think that P presentation is more coherent than Einstein presentation because it is well known that classical rigidity (instantaneous action-at-a-distance, à la Descartes) is incompatible with Einstein's SR (ESR) as well than Poincaré's SR (PSR). *But the important concept according the spirit of the text is not the rigidity but the identity:*

Let there be given a stationary *rigid rod*; and let its length be L as measured by a measuring-rod which is also stationary. (…)
*In accordance with the principle of relativity* (…) « *the length of the rod in the moving system* » - must be equal to « *the length L of the stationary rod.* » (...)
The length to be discovered [by LT] we will call « *the length of the (moving) rod in the stationary system*». This we shall determine on the basis of our two principles, and we shall find that it differs from L. (My italics but Einstein's quotes, 3, §2,1905).

Max Born, who was a specialist of rigidity in Einstein's special relativity wrote in 1921 in his book on relativity that Einstein introduces a tacit assumption:

A fixed rod that is at rest in the system S and is of length 1 cm, will, of course, also have the length 1 cm, when it is at rest in the system S', provided that the remaining physical conditions are the same in S' as in S. Exactly the same would be postulated of the clocks. We may call this *tacit assumption* of Einstein's theory the *"principle of the physical identity of the units of measure"*. [14, my italics, p252]

This is not a third hypothesis because Einstein's deduces the identity of his rods from his relativity principle[9]. The rigidity is not important. The important thing in the spirit of the young Einstein's text, is to postulate the existence in Nature of processes giving units of length and time.

Each SR rest on its own system of axioms:

---

[8] According to A. Pais [16], the "third" Poincaré's hypothesis proves that Poincaré has not understood the SR. Pais doesn't try to penetrate the logic of Poincaré's relativity.

| **Poincaré's principles of SR** | **Einstein's principles of SR** |
|---|---|
| (implicit kinematics of deformable rods) | (explicit kinematics of rigid rods) |
| 1 principle of relativity (compensation) | 1 principle of relativity (identity) |
| 2 principle of real contraction (lengths and units, LH). | 2 principle of the invariance of (the one way) speed of light (Light principle, LP) |

## 3 Poincaré's use and Einstein's use of LT

It is well known that the compatibility with the null-results of the Michelson's experiment is based on real contraction of lengths in Lorentz-Poincaré's point of view.

We don't want discuss here the origin of length contraction or more exactly "the hypothesis FitGerald-Lorentz deformation" that is a very difficult problem. It is often claimed that this original contraction is purely longitudinal and that it is an "ad hoc" conjecture. Lorentz's demonstration $l=1$ for the scale factor was however very arguable and the nature "ad hoc" conjecture have deep physical foundations in the atomic structure in Lorentz's theory [15-3].

We want here to concentrate the attention on Poincaré representation of LH.

Firstly, if in 1900 Poincaré underlines the nature "ad hoc" of LH it is only because the local time and the real contraction are not *connected* in Lorentz's theory.

Secondly, if Poincaré insists on his own contribution, the demonstration $l(\varepsilon)=1$, it is clearly because he considers a purely longitudinal Lorentz's contraction [21] in order to show the compatibility of this latter with principle of relativity [20-2]:

So Lorentz hypothesis [LH] is the only one that is compatible with the impossibility of demonstrating the absolute motion [RP]; if we admit this impossibility, we must admit that moving electrons are contracted such a manner to become revolution ellipsoids whose two axis remain constant. [9b, §7]

This last sentence is particularly interesting because LH is not a consequence of principle of relativity but it is an independent hypothesis. Poincaré admits - as Lorentz - that the

---

[9] Identity of the units can be also deduced from Einstein's second principle, good understood as the strictly numerical identical value of the speed of light within each system K and k.

contraction is real but - contrary to Lorentz - he *raises* this Lorentz's hypothesis - justified on the basis of the atomic structure of the matter- *to a status of a postulate (LH is the only one compatible with RP)*. We must understand why Poincaré's two principles are as compatible than Einstein's two principles. Poincaré writes, in his sixth § of his 1905 work, on LH:

In accordance with LH, moving electrons are deformed in such a manner that the real electron becomes an ellipsoid, while the ideal electron at rest is always a sphere of radius r (…) The LT replaces thus a moving real electron by a motionless ideal electron. [9b, §6]

Poincaré's *compatibility LH and RP* implies another use of LT than the standard one. In order to illustrate this, we can use Tonnelat's diagram [22] (fig 1, *we adopt Poincaré's and Einstein's respective notations* in the following, see former respective quotations about LT):

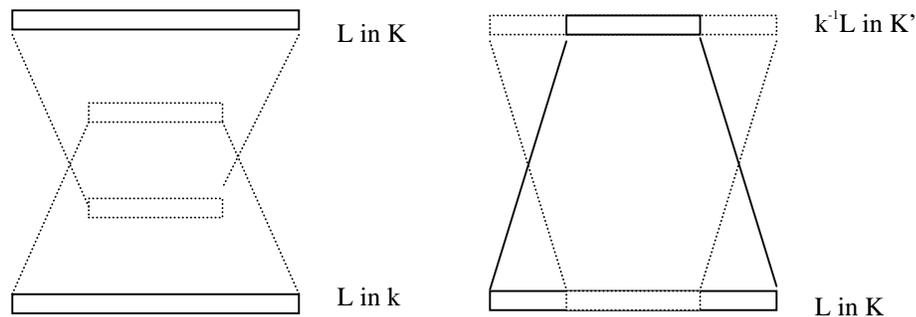

**Fig 1**
In Einstein's words
(dashed lines):
"The length to be discovered [with LT]"the length of the (moving) rod in the stationary system"

**Fig 2**
In Poincaré's words
(dashed lines):
"LT replaces thus a moving real electron [rod] by a motionless ideal electron [rod]"

In **fig 1**, in Einstein's standard SR, the contraction of the moving rod $\gamma^{-1}L$ is not real (dashed line) but is the reciprocal result of a comparison of measurement made on identical rods L (continuous lines) from one system to the other with the well known use of LT (dashed lines).

In **fig2**, in Poincaré's SR the contraction of the moving rod $k^{-1}L$ is *by principle* (LH) real (continuous lines) in K'. By the use of LT the length of the rod in K' (for observers in K') seems be equal to L (dashed lines). *Reciprocally*, we can of course reverse the role of K and K' (where ether is now chosen at rest) and reverse the continuous lines and dashed lines. Contrary to Lorentz's point of view, the contraction is reciprocal because of course we can always choose the system in which the ether is at rest.

The calculation with the LT is also very easy. Suppose the ether is chosen by definition at rest in K. The real length of the rod placed in the moving system K' is thus $k^{-1} L$. The first LT $x' = k (x - \varepsilon t)$ "replaces" (in Poincaré's terms) in any time t (see below) the length moving real rod $k^{-1}L$ by a motionless rod L.

Poincaré's principles are as compatible than Einstein's principles but the historical difference is that Poincaré has never developed *explicitly* his way of use of LT on a basic exemple (a deformable rod). According to Poincaré's implicit kinematics, the real differences are *compensated* by a "good use" of LT. According to Einstein's explicit kinematics, the *identical* processes seem to be different by another "good use" of LT. This is of course valid for the rods but also for the clocks.

It is also particularly important to notice that Poincaré's conception of real contraction of units of measurement (observers K' measure *really* shorter lengths with *really* shorter units of length) is in perfect harmony with LT:

The contraction is the same for all bodies: then, how could we perceive it? Then, measure cannot give us any information because the *metre*, put in the sense of the movement, is also contracted. [11,1908]

There exists therefore a genuine antinomy (in a Kantian sense) between Einstein's conception of and Poincaré's conception of units. It would be however a serious mistake to believe that the difference between the two SR would be *only* metaphysics (definition of realities and appearances).

We shall show that the difference is deeply physics in the sense that Poincaré's use of LT is based on a classical conception and Einstein's use of LT on a quantum conception of units.

## 4 Poincaré's and Einstein's conventions of synchronisation

In 1911 Poincaré is aware that the German relativistic school (Einstein, Planck, Laue, Sommerfeld, Minkowksi…) has adopted another assumption for the SR ("une autre convention"):

Today some physicists want to adopt a new convention. This is not that they have to do it; they consider that this convention is more easy, that's all; and those who have another opinion may legitimately keep the old assumption in order not to disturb their old habits. [12]

Poincaré and Einstein use the same method of distant clock synchronisation (exchange - forth and back - of signals of light) but we shall show that Poincaré's convention is not the same as Einstein's convention.

The synchronisation method by exchange of signals of light is developed by Poincaré in 1900 in a paper [7] on the reaction principle in the Lorentz theory. Poincaré explains that if the Lorentz's local time $t' = x - vx/c^2$ is used in the system K' moving with respect to the ether, the observers remark no difference (*at first order*) between the forth travel time and the back travel time of the light. For the second order Poincaré envisages already in 1900 that the hypothesis of Lorentz is necessary.

If it is true that Poincaré's second principle has the same statute than Einstein's second principle, we must show that Poincaré's conception of time can be deduce from LH exactly in the same manner that Einstein's conception of time is deduce from LP (with principle of relativity of course in the two cases).

Once again Lorentz's conception of local time ($t' = t - \varepsilon x$) is the starting point for Poincaré[10] (1900) but not the final point in Poincaré's SR.

In others terms both independent (and thus ad hoc according to Poincaré) Lorentz's hypothesis (real contraction and local time) have to be connected in a as harmonious way than Einstein's one.

In his talk on "The principles of the mathematical physics", Poincaré explains (we break down into two parts his argumentation):

**1-** (Two observers are at rest relative to ether, System K)

The most ingenious idea has been that of local time. Imagine two observers [A and B] who wish to adjust their watches by optical signals; they exchange signals, (…) And in fact, they [The clocks of A and B] mark the same hour at the same physical instant, but on one condition, namely, that the stations are fixed.

**2-** (Two observers are moving relative to ether, System K')
In the contrary case the duration of the transmission will not be the same in the two senses, since the station A, for example, moves forward to meet the optical perturbation emanating from B, while the station B flies away before the perturbation emanating from A.
The watches adjusted in that manner do not mark, therefore the *true time*; they mark the *local time*, so that one of them goes slow on the other (de telle manière que l'une retarde sur l'autre). It matters little, since we have no means of perceiving it. (…)
(for exact compensation, we must add LH, former quotation)
Unhappily, that does not suffice, and complementary hypotheses are necessary. It is necessary to admit that bodies in motion undergo a uniform contraction in the sense of the motion. [7]

Poincaré's synchronisation is clearly based for his second system K' on the *duality of the true time t and the local time t'*.

This is Poincaré's "tour de force" to have shown that his second principle (LH) implies for the local time the expression given by the fourth LT: t' = k (t – εx). But it is also his weakness because his complete historical demonstration in his lectures in La Sorbonne, based on lengthened light ellipsoids, is complicated.

A direct calculation with LT is however possible. If the two observers A and B (fixed in K'), who exchange light signals, are in movement relatively to K, the forth travel in *true time* is not the same than the back travel in true time. Indeed for Poincaré the light is a wave (in ether) whose speed is independent of the speed of the source but is not independent of the speed of the frame K'.

Let us introduce the second principle LH: $k^{-1}L$ is the real distance between the two observers fixed in K'. By using the fourth LT, the *local time* t' = k (t – εx), it is easy but to see that the difference is exactly *compensated because the apparent distance AB is L for the observers in K'* and everything happens as if the speed of light was the same forth and back in K'. Everything happens also as if Poincaré's use of LT brings the ether back to rest in K'

In summary we have in Poincaré's logic:

$$\boxed{\text{LT (principle of relativity) + LH (real contraction)} \rightarrow \text{Local time (fourth LT)}}$$

Now we haven't yet answered the fundamental question: why are both conventions of synchronisation deeply different?

Indeed Poincaré adopts in his system K where the ether is chosen at rest exactly the same convention (assumption) than Einstein in his stationary system K:

**1-** (Einstein's "stationary time of a stationary system K")

But it is not possible without further *assumption* to compare, in respect to time, an event at A with an event at B. We have so far defined only an "A time" and a "B time". We have not defined a common "time" for A and B, for the latter cannot be defined at all unless to establish *by definition* that the time it required by light to travel from A to B equals the time it requires to travel from B to A.

Let a ray light start at the "A time $t_A$" from A towards B, let it at the "B time" $t_B$ be reflected at B in the direction of A, and arrive again at A at the A time $t'_A$. In accordance with definition the two clocks synchronize if

$$t_B - t_A = t'_A - t_B.$$

---

[10] Lorentz's local time is also an ad hoc assumption (it is a mathematical change of variable, [1]).

(…) It is essential to have time defined by means of stationary clocks in stationary system (…) [3, §1, 1905]

But what happens for Einstein's second system k? The repetition of the concept *stationary* is essential because in his third §, Einstein notices [10] about his second system k (ξ, η, ζ, τ).

**2-** (Einstein's "stationary time of a stationary system k")
To do this[11] [deduce LT] we have to express in equations that τ is nothing else than the set of data of clocks at rest in system k, which have been synchronized according to the rule given in paragraph 1.[3, §3, 1905]

The synchronisation of identical clocks within the second system k is exactly the same than the synchronisation in the first system because the speed of light is of course exactly the same. In Einstein's own terms "as demanded by the principle of relativity and the constancy of the speed of light also propagates with velocity c in the moving system".

Einstein's radical elimination of ether implies that his two systems, K(t) and k(τ), are prepared in internal identical states of synchronisation. This is the reason for which the duality true time-local time have no sense in Einstein's logic. Poincaré's relativistic ether is always at rest in a given inertial frame (his relativistic ether hasn't got a particular state of movement), but *the other one is then moving relative to the ether*.

In Poincaré's logic, the clocks in K' are adjusted in that manner the real difference of duration in true time is (exactly with LH) compensated *after* the use of LT and therefore the local time (fourth LT). Poincaré notices in 1904 (we repeat the former quotation):

The watches adjusted in that manner do not mark, therefore the true time; they mark the local time, so that one of them goes slow on the other (de telle manière que l'une retarde sur l'autre). It matters little, since we have no means of perceiving it. [7]

English translation ("goes slow") is perhaps ambiguous: Poincaré means that K' clocks are not put in the same origin of time. There are adjusted with respect to the true time but they are not adjusted a priori with each other. There is the deep difference between the two synchronisation because Einstein's preparation of stationary systems where the k clocks are synchronised which each other - before the use of LT – provides a truly identical rhythm or rate of the clocks.

---

[11] Einstein admits a priori the same relation of synchronisation within the two systems and he deduces LT (or more exactly LT of the co-ordinate of an event) from this relation [3, §3].

It is indeed easy to see that Einstein's synchronisation and Einstein's identity of units are exactly the same concept. With identical rods and with the (one-way) speed of light c numerically identical within the two systems, we have of course the same internal duration. Einstein's concept of "synchronous clocks" is very subtle because the repetition of the process of synchronisation (sometimes called Einstein's light clock) provides an identical rhythm (unit of time) for the clocks.

We can now deeply understand why Poincaré never speaks of duration given by identical clocks or a fortiori never speaks of dilation of such a duration (implicit Poincaré's dilation [20, p154] is of course also real but it is a consequence of his second principle LH).

Poincaré's local time is not a *internal time* in the second frame K'. It is only an "auxiliary time" or a "dependent time" with respect to true time. More deeply the local time t', is defined as a func*tion* of the independent t the true time[12]. Independent doesn't mean absolute – as in Lorentz's theory - but only that t is a parameter. Poincaré's temporality is as physical than Einstein's one but it belongs to classical physics ("La Mecanique Nouvelle", with his relativistic kinematics and relativistic dynamics).

 **5 Einstein's preparation of identical isolated stationary systems and the adiabatical hypothesis**

There is another way to prove the existence of a "structure fine" of SR. Indeed, the young Einstein distinguishes, in original 1905 and 1907[13] articles two stages in his preparation of inertial frames:

*first stage. The preparation of the two systems in state of rest:*

---

[12] We have shown that Einstein's concept of event (x, y, z, t) and Poincaré's concept of local time (x (t), y (t), z(t), t'(t) are antinomic (in a Kantian sense).
[13] Einstein's formulation of 1905 is of course the same: "Let each system be provided with a rigid-measuring-rod and a number of clocks, and let the two measuring-rods, and likewise all the clocks of the two systems, be in all respects alike"

"Let us consider K et k two equivalent systems of reference; we may say that the systems have measuring-rod of same length and clocks giving the same indications, the comparison between this objects being made when they are *in state of relative rest* (im Zustande relative Ruhe miteinander) " [4, §1-1907]

*Second stage[14]: the "launching of the boost":*

"Now let a constant velocity v (Es werde nun dem Anfangspunkte … erteilt) to the origin of one of two systems (k)" [3, §1-1905]

The young specialist of statistical thermodynamics Einstein formulates explicitly in 1907 the hypothesis (I discuss this adiabatical hypothesis[15] in my book) that his identical clocks and his "rigid" rods are not modified by the passage from the velocity to the velocity v.

I don't insist on this point here but only on the fact that a *finite time* is necessary to synchronise the two *distant* clocks when they are at rest (Poincaré never puts his frames at rest during a finite time).

Many authors think that the problem of rigidity in young Einstein's text can only be solved in general relativity (GR) because the concept of rigidity is deeply connected with Euclidean geometry.

We propose another way of research. If Einstein's original preparation of his isolated stationary frames (with his adiabatical hypotheses[16]) at rest is forgotten, the internal state of the second system remains identical when it returns to another non-accelerated state. We must then introduce a truly *quantum of time* (an identical unit within both frames) in Einstein's logic. This is a quantum solution of the problem of rigidity in Einstein's SR (without GR). The idea of use light signals to define distances in Einstein's SR (Bondi H., the factor k) is necessary but not sufficient. We absolutely need of a quantum of time (identical unit within the two systems).

---

[14] There is of course a well known *third stage* in Einstein's logic (his paragraph 4): when the system k is in moving relative to the system K, it is impossible for an observer in K to compare directly the rigid rods or the clocks of his system with the rods and the clocks of the other system k. He *has to use* LT to rely the two systems. The dilation of time as contraction of length is not real but is the (reciprocal) result of a comparison of measurement made from one system to the other.

[15] Van der Waerden reminds of the crucial importance of the adiabatical hypothesis formulated explicitly by Ehrenfest in the development of quantum theory. Two important heuristic principles have guided quantum physicists during the period 1913-1925. Ehrenfest's adiabatic hypothesis and Bohr's principle of correspondance. The adiabatic hypothesis, first formulated by Ehrenfest in 1913 ("A theorem of Boltzmann and its connection with the theory of quanta"): if a system be affected in a reversible adiabatic way, allowed motions are transformed into allowed motions. The name adiabatic hypothesis is due to Einstein as Ehrenfest states in his paper. Finally, Ehrenfest shows that the adiabatic hypothesis is closely connected with the second law of thermodynamics. I showed in my book the crucial importance of the invariance of entropy in Einstein's SR.

Poincaré's temporality and Einstein's temporality are both relativistic but the time is not a parameter in Einstein's SR. In Einstein's own words, time (t or $\tau$) is a *set of data of identical clocks* (Gleichbesshaffene Uhren) that beat time with identical rhythm.

*This set is infinite but countable.* Poincaré's set of points of true time is also infinite but non-countable [20, p262]. The development of both logics lead us to a very interesting mathematical contrast between "countable set" and "non-countable set" [17].

Einstein's " *set of data of clocks"* $\tau$ represents essentially a set of values or a spectrum of values of the observable time. Einstein's internal time called proper time –eigenzeit- by Minkowski in 1908 is essentially a duration of time (Zeitteilchen, Zeitelement in Einstein's 1905 words). When Poincaré's hidden variable (see annex 1), the true time, is radically eliminated in the preparation of the second system (K' -> k or t' -> $\tau$ ), we may say that the set of values of the "eigenzeit" is exactly the same concept than the set of eigenvalues[18] of the observable "Zeit".

## 6 Einstein's quantum clocks and the spectral identity of atoms

We propose now another demonstration of the quantum nature of time in Einstein's SR on the basis of an analysis of instruments of measure of the observable time. The young Einstein identifies explicitly "clock and atom" in his second fundamental synthesis on SR in 1907:

Since the oscillatory process that corresponds to a spectral line is to be considered as a intra-atomic process, whose frequency $\nu$ is determined by the ion alone, we can consider such an ion as a clock of definite frequency $\nu_0$; this frequency is given, for example, by the light emitted by *identically* constituted ions at rest with respect to the observer. [4, §3].

The young specialist of statistical thermodynamics is frighteningly clear-sighted. His intuition implies not only that the atoms ("producers of spectral lines" in Einstein's own terms) of a same nature are identical but also that this identity permit us to known the frequency in his own system (after Minkowski, we say proper frequency) if the atom (the clock) is moving relative to us (the observers).

---

[16] We have shown [20-2] that Einstein's adiabatical hypotheses for his isolated stationary systems is deeply connected with Ehrenfest's adiabatical hypotheses whose essential role in the beginning of quantum theory is well known.

[17] We arrive at the same conclusion from an analysis of deeply discontinuous Einstein's concept of independent events [20-1].

[18] My research on the quantum interpretation of Einstein's synchronisation is also the starting point of S. Reynaud's and T. Jaeckel's research (Jussieu) [16] on time operator and on the statute of acceleration in SR (without GR):"The physical observables describing space-time positions cannot be confused with classical co-ordinate parameter (…). Their definition has to reach limits associated with quantum nature of the physical world."

There exist therefore in Nature physical process giving identical units of time *within each inertial system*. In the framework of classical mechanics, the *spectral identity* of atoms is not comprehensible. Weiskopf underlines this essential point – often forgotten – of the quantum conception:

The main idea of quantum theory, I said, there is idea of identity. (…) Understanding the idea of identity, there is the understanding the concept of quantum state established by Bohr in the first period of his scientific activity. [24-1]

Within the framework of prequantum concepts two objects could not be *identical* in every respect since, in principle prequantum physics requires an infinite set of indications for the full description of an object. It could always differ in some very small detail. The orbit of an electron around the nucleus differs by some amount. Indeed it would be extremely improbable to find two atoms with exactly the same electrons orbit. Therefore a new conceptual framework was needed in which the state of a system is fully define in all his qualities by a finite set of indicators. This new framework was quantum mechanics and its leading concept is quantum state.[24-2]

In order to have his identical units of measure, the young Einstein requires not only the classical concept of identity (in the sense of orbital identity in classical statistical mechanic or in the sense of minimum scale of classical chemistry) but the quantum concept of spectral identity. The identical processes must be associated to identical units of time (or a frequency of course).

My argument is not only a argument based on historical foundations. Indeed, as the unit of time is identified by Einstein ($\tau = 1/\nu$) to the (inverse of) frequency (of the spectrum), we absolutely need a purely monochromatic wave of a determined frequency. Einstein's intuition of identity ($\tau = 1/\nu$) can only be justified in the framework of quantum theory because the emission of monochromatic $\nu$ radiation is directly connected to the concept of quantum state of the atom ($E = h\nu$).

Moreover the most radical conception that identifies thermodynamically a monochromatic radiation with a set of independent quanta of light is of course his own conception:

"From this we further conclude that monochromatic radiation of low density behaves thermodynamically as if it consisted of mutually independent energy quanta"[2, §6]

It is unquestionably specifically Einsteinian. Now if Einstein's clocks are quantum clocks, what are Poincaré's clocks? In "La mesure du temps" [2], The great specialist of Celestial Mechanics writes in 1998:

"In fact the best clocks have to be corrected from time to time, and corrections are made with the astronomical observations; (…) In other terms, it is the sidereal day or the duration of rotation of the Earth, that is the constant unit of time. [6]

It is sometimes claimed that Poincaré's analysis of distant simultaneity (1898) is the same than Einstein's one, seven years later. But it is completely wrong because Poincaré, in "La Mesure du temps", underlines the conventionality of simultaneity[19].

Moreover we proved that Einstein's definition of distant simultaneity is exactly the same thing than Einstein's principle of identity of units. In the same text Poincaré writes:

"When we use the pendulum to measure time, which is the postulate that we admit implicitly? It is that duration of two identical phenomena is the same. Watch out one moment (Prenons-y garde un instant). Is it possible that experiment denies our postulate.? If experiment made us the observers of such a spectacle our postulate would be contradicted." [6]

In a paper in 1910 on his SR Einstein affirms explicitly his postulate:

"Thus, we postulate that two identical phenomena are of the same duration. The perfect clock so defined plays a role in the measurement of time that is analogous to the role played by the perfect solid in the measurement of lengths."[5]

The borderline classical-quantum passes clearly between the two SR and the existence of a "fine structure" of SR [20-2] is thus established on foundations that are non only metaphysical but physical: There is a SR with Einstein's quantum clocks and SR with Poincaré's classical clocks.

---

[19] In my opinion, Poincaré's analysis of conventionality of simultaneity is the same than Reichenbach's analysis 23 years later. We show in annex the connection between the famous Reichenbach's parameter and Poincaré' synchronisation.

In order to take a well known Galileo's expression we might say that there is a SR from the messenger of the atoms who is Einstein and a SR from the messenger of the stars who is Poincaré.

**Conclusion: standard mixing SR and metrical structure of space time**

The idea to try to underscore the quantum features of the metrical structure of space-time is of course not a new idea. So Brown writes:

"Anandan follows a suggestion of Penrose and argues for the quantum mechanical nature of clocks in their fondamental role as hodometers of the metrical structure of space-time.
I see no reason why a version of this thesis should not apply also to Galilean-Newtonian and Minkowski space-time. In the present state of research, this thesis is not demonstrated."[15-1, §IX]

I think that H. Brown is right and that this thesis is not yet demonstrated essentially because it is impossible to solve the problem directly from a analysis of Minkowski's metric. Indeed, if both conceptions of Einstein and Poincaré are already not very easy to distinguish, it is still much more difficult to distinguish Minkowski's and Poincaré's approachs of space-time.

SR is today in a state of mixing and is particularly true for standard Minkowski's geometrical representation of SR. I showed [20-3] that it is possible to separate Poincaré's four-dimensional and Minkowski's four-dimensional representation and thus to give a geometrical sense to Einstein's principle of identity of units of measure.

In this paper I didn't analyse the geometrical concept of Minkowski's metric and I only showed that Penrose's suggestion (quantum nature of clocks) is true for one of the two components of the mixing: Einstein's SR.

This is essential for the opportunity of quantization in General Relativity as it is noticed by Brown if we assume (it is not sure) that Einstein's SR is the local limit of Einstein's GR

"The question is raised as to whether it is correct to consider the metric field as an entity that should be quantised. It is in a sense, already an entity with quantum mechanical pedigree."[15-2]

19/11/00 (Yves Pierseaux)

**Appendix A: covariance of the speed of light deduced from Poincaré's principles**

Let us thus examine Poincaré's convention of synchronisation. Consider two systems of reference K (ether) and K' where there are two stations A and B placed in the direction of the movement (velocity v).

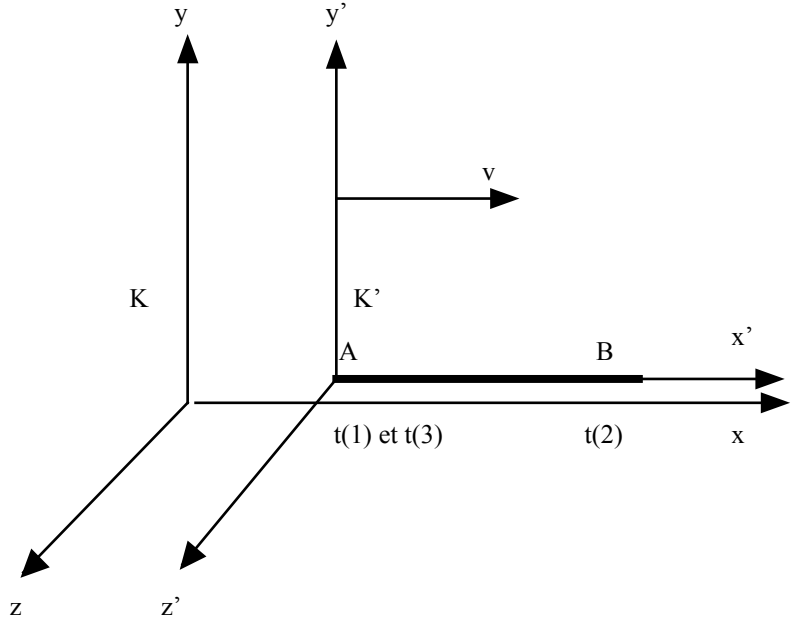

Let start an electromagnetic wave with the velocity c (in the ether, $\varepsilon = v/c$) from A towards B (1), let it be reflected at B in the direction of A (2), and arrive again at A. The length *really contracted* AB of the *deformable rod* is $k^{-1} L$.

We shall estimate thus the forth an back travel time (from A to B and from B to A) of the electromagnetic wave successively *in true time* t and in *local time* t' (3).

<u>1) FORTH AND BACK TIMES TRAVEL IN TRUE TIME</u>

Suppose that A is the common origin of the two systems at time $t(1) = 0$.

We have thus $x_A(1) = 0$ et $x_B(1) = k^{-1} L$.

We shall seek the co-ordinates of arrival of the wave at A *in true time* by the evaluation of the distance covered by the light in the system K (with the speed c).

Let $t(2)$ the time for the electromagnetic wave to arrive in B. During this time the point B is translated towards the right of $v\, t(2)$. The distance covered by the light is thus $c\, t(2) = k^{-1} L + v\, t(2)$ We have ($\varepsilon = v/c$):

$$c\, t(2) = L\sqrt{1 - \varepsilon^2} + v\, t(2) \quad \text{and thus} \quad t(2)(1 - \varepsilon) = \frac{L}{c}\sqrt{1 - \varepsilon^2}$$

So we obtain the time t(2) of arrival in B:

$$t(2) = \frac{L}{c}\sqrt{\frac{1+\varepsilon}{1-\varepsilon}}$$

The co-ordinate $x_B(2)$ is

$$x_B(2) = v\frac{L}{c}\sqrt{\frac{1+\varepsilon}{1-\varepsilon}} + \sqrt{1-\varepsilon^2}\,L = \varepsilon L\sqrt{\frac{1+\varepsilon}{1-\varepsilon}} + \sqrt{1-\varepsilon^2}\,L$$

Now let the time t(3) for the light to return in A. The distance covered by the light, c t(3), is the sum of the forth distance covered by the light during the time t(2) and of the back distance covered by the light during the time t(3) - t(2):

$$c\,t(3) = [L\sqrt{1-\varepsilon^2} + v\,t(2)] + [L\sqrt{1-\varepsilon^2} - v(t(3)-t(2))]$$

We obtain for the time t(3):

$$t(3) = 2\frac{L}{c}\frac{1}{\sqrt{1-\varepsilon^2}}$$

The co-ordinate $x_A(3)$ is:

$$x_A(3) = v\,t(3) = v\,2\frac{L}{c}\frac{1}{\sqrt{1-\varepsilon^2}} = 2\varepsilon L\frac{1}{\sqrt{1-\varepsilon^2}}$$

So we have

$$\text{FORTH: } t(2) = \frac{L}{c}\sqrt{\frac{1+\varepsilon}{1-\varepsilon}} \qquad \text{BACK: } t(3)-t(2) = 2\frac{L}{c}\frac{1}{\sqrt{1-\varepsilon^2}} - \frac{L}{c}\sqrt{\frac{1+\varepsilon}{1-\varepsilon}}$$

In Poincaré's SR with ether, the equality of the back-and-forth travel times is not postulated *in true time*. The true time determines the real state (of synchronization) of the primed system K'.

## 2) FORTH AND BACK TRAVEL TIMES IN LOCAL TIMES OF A AND B

In the SR with ether the LT defines the local time (at A or at B):

$$t'_A = k(t - \frac{v}{c^2}x_A) \qquad t'_B = k(t - \frac{v}{c^2}x_B)$$

Let $t'_A(1)$, $t'_B(2)$ and $t'_A(3)$ the locals times of the departure, arrival and return on the light's wave. We have

$$t'_A(1) = k(t(1) - \frac{v}{c^2}x_A(1)) \qquad t'_B(2) = k(t(2) - \frac{v}{c^2}x_B(2)) \qquad t'_A(3) = k(t(3) - \frac{v}{c^2}x_A(3))$$

From the first equation, we have $t'_A(1) = 0$.

From the second equation, we replace t(2) et $x_B(2)$ by their values and we find:

$$t'_B(2) = k\left(\frac{L}{c}\sqrt{\frac{1+\varepsilon}{1-\varepsilon}} - \varepsilon^2 \frac{L}{c}\sqrt{\frac{1+\varepsilon}{1-\varepsilon}} - \varepsilon \frac{L}{c}\sqrt{1-\varepsilon^2}\right)$$

$$t'_B(2) = \frac{L}{c}$$

The forth travel time $t'_B(2) - t'_A(1)$ is thus $L/c$.

From the third equation, we replace t(3) et $x_A(3)$ by their values, and we find:

$$t'_A(3) = k\left(2\frac{L}{c}\frac{1}{\sqrt{1-\varepsilon^2}} - 2\varepsilon L \frac{1}{\sqrt{1-\varepsilon^2}}\right)$$

We obtain:

$$t'_A(3) = 2\frac{L}{c}$$

The back travel time, $t'_A(3) - t'_B(2)$ equals also to $L/c$.

It is so that the covariance of the speed of light is deduced from the two Poincaré's principles ("local time" and "real contraction").

## Appendix B: Reichenbach's parameter and the ε-LT of Poincaré

Reichenbach's aim [22] is, in the starting point, the same than Poincaré's one in 1898 [6]: the demonstration of the conventionality of the distant simultaneity or more precisely the comparison of the travel forth and the travel back of a light signal in a given inertial system.

Reichenbach has introduced a parameter $\kappa$ (I adopt the letter $\kappa$ for Reichenbach's parameter in place of the standard notation $\varepsilon$ to avoid the confusion with Poincaré's notation $\varepsilon$) whose value is the travel-time forth divided by the travel-time forth-back.

It is elementary to evaluate this parameter in both SR.

**Einstein's SR**

The "one-way speed of light"[25] is of course an invariant in the relativistic logic of Einstein. It is well known that this parameter equal to one half *within* all inertial systems in SR without ether. In particular we have in each Einstein's frame (K and k): $\kappa = ½$ for

**Poincaré's SR**

For Poincaré's system K, $\kappa = \frac{1}{2}$ because ether is by definition at rest in this system.

For Poincaré's system K', we must distinguish the situation in true time and in local time (with LT)

**In true time t**, we have calculated (see annex 1)

FORTH: $t(2) = \frac{L}{c}\sqrt{\frac{1+\varepsilon}{1-\varepsilon}}$    BACK: $t(3) - t(2) = 2\frac{L}{c}\frac{1}{\sqrt{1-\varepsilon^2}} - \frac{L}{c}\sqrt{\frac{1+\varepsilon}{1-\varepsilon}}$

The value of Reichenbach's parameter $\kappa$, may be determined in function of the speed $\varepsilon$ relatively to the ether:

$$\kappa = \frac{\sqrt{\frac{1+\varepsilon}{1-\varepsilon}}}{\sqrt{\frac{1+\varepsilon}{1-\varepsilon}} + \sqrt{\frac{1-\varepsilon}{1+\varepsilon}}} = \frac{1+\varepsilon}{2}$$

The connection between Reichenbach's parameter and the relative speed $\varepsilon$ with respect to the ether is quite natural.

$$\kappa \neq 1/2 \text{ in K' in true time.}$$

We have only $\kappa = 1/2$ when $\varepsilon$, the speed relative to the ether, equals to zero. This result seems trivial but precisely we shall show that the elimination of the ether by Einstein may be *interpreted as if* the two systems, K and k, were in the same state (at rest, $\varepsilon = 0$) of synchronization.

**In local time t' (with LT)**, we have calculated (see annex 1):

"The back travel time, $t'_A(3) - t'_B(2)$ equals also to L/c."

and thus $\kappa = \frac{1}{2}$ in K' in local time

*All happens as if* when A and B used their local time, the forth travel time was the same as the back travel time. But the two times are different in the "reality".

In my opinion, Poincaré's convention of synchronisation is the definitive *physical* answer to the problem of Reichenbach's parameter. Indeed if we consider only one inertial system we can discuss without end on the value of Reichenbach's parameter. The physics begins when it is possible to compare experiments from two systems in uniform translation (this is not only true for Poincaré and Einstein but also for Galilee). The value of the parameter, in true time, for the second system K' is $\kappa \neq \frac{1}{2}$ that becomes $\kappa = 1/2$ in local time.

It is therefore a mistake to introduce this parameter (or another synchronisation's parameter) in LT. *According to the "fine structure" of SR, The κ-Lorentz transformation is in fact the ε-Lorentz transformations of Poincaré.* I showed in my paper "Euclidean Poincaré's SR and non Euclidean Einstein-Minkowski's SR" that the genuine physical contrast is situated between Einstein's **v**-LT of independent events and Poincaré's ε-LT of the coordinates of a material point.

# References
## primary sources